\documentclass[journal,twoside,web]{ieeecolor}
\usepackage{generic}
\usepackage{cite}
\usepackage{amsmath,amssymb,amsfonts}
\usepackage{graphicx}
\usepackage{textcomp}
\usepackage{epstopdf}
\usepackage{siunitx}
\usepackage{amssymb}
\usepackage{subfig}
\def\BibTeX{{\rm B\kern-.05em{\sc i\kern-.025em b}\kern-.08em
    T\kern-.1667em\lower.7ex\hbox{E}\kern-.125emX}}
\begin{document}

\title{Modeling Interface Charge Traps in Junctionless FETs, Including Temperature Effects}

\author{Amin Rassekh, Farzan Jazaeri, Morteza Fathipour, and Jean-Michel Sallese
\thanks{Amin Rassekh, Farzan Jazaeri, and Jean-Michel Sallese are with the Electron Device Modeling and Technology Laboratory (EDLAB) of the Ecole Polytechnique F\'{e}d\'{e}rale de Lausanne, Switzerland (email: amin.rassekh@epfl.ch). Amin Rassekh and Morteza Fathipour are with device simulation and modeling Laboratory, department of electrical and computer engineering, University of Tehran, Iran. Corresponding authors: Morteza Fathipour, email: mfathi@ut.ac.ir and Amin Rassekh, email: rassekh.amin@ut.ac.ir. received XXXX XX, 2019.}\vspace{-1cm}}
\maketitle

\begin{abstract}
In this paper, an analytical predictive model of interface charge traps in symmetric long channel double-gate junctionless transistors is proposed based on a charge-based model. Interface charge traps arising from the exposure to chemicals, high-energy ionizing radiation or aging mechanism could degrade the charge-voltage characteristics. The model is predictive in a range of temperature from 77K to 400K. The validity of the approach is confirmed by extensive comparisons with numerical TCAD simulations in all regions of operation from deep depletion to accumulation and linear to saturation.
\end{abstract}


\section{Introduction}
\IEEEPARstart{J}{unctionless} field-effect transistors (JLFET) were proposed as a promising device for the future of scaling. Actually, source/drain (S/D) junctions in conventional metal-oxide-semiconductor field-effect transistors (MOSFET) is a big challenge in the nanoscale dimension \cite{lee2009junctionless}. The main advantage of JLTs is that they overcome this step and still are fully compatible with CMOS technology. In addition, JLFET is a good option for label-free biosensors when shoped as nanowires \cite{cui2001nanowire}.

Using JLFET as biosensors, called Nanowires (NWs), is an interesting option due to their proven high sensitivity to biological and chemical elements \cite{yesayan2016charge}. They share the same principle of operation with regular JLFETs, except that they do not have any solid-state gate over the insulator. Because of this feature, they are  exposed to contaminants from the materials which they are in contact with. Hence, these devices are facing much more defects or interface charge traps than the state-of-the-art CMOS devices, a situation that may affect their electrical properties. Similarly, trap and defects can be generated by ionizing radiation just as in regular MOSFETs.

When a MOSFET is exposed to high-energy ionizing radiation, deep trap states are generated in the bulk oxide or near the Si/oxide interface. The process of traps generation depends on the temperature, the applied electric field, and the oxide thickness \cite{pezzotta2016impact} and \cite{zhang2016gigarad}. Aging mechanism in MOS technologies is also a common cause of traps creation at the Si/oxide interface \cite{huard2009cmos} and \cite{esqueda2015compact}. These interface traps could degrade the performance of MOS devices.

Modeling the effect of radiation on inversion mode MOSFETs has already addressed in \cite{jazaeri2017charge}, but that model is not appropriate for junctionless devices. Influence of interface charge traps on biosensor junctionless devices was discussed in \cite{yesayan2016charge}.

Thus, in order to account for ionizing radiation, stress-induced defects as well as the effect of interface charge traps on JLFETs in a simple and compact model compatible approach, we propose analytical expressions for modeling the interface charge traps in Double-Gate JLFET (DG JLFET), and we also include the effect of temperature from \SI{77}{K} to \SI{400}{K}. This model relies on the charge-based approach in \cite{sallese2011charge} where we derive an equivalent gate-source voltage ($ \Delta V_{gs} $) to take into account the effect of interface charge traps. This is quite different from former developments \cite{yesayan2016charge}. We study the behavior of current-voltage characteristics and derive analytical solution for the subthreshold swing (SS). This approach will be validated with technology computer-aided design (TCAD) simulations, including temperature.

\section{GATE VOLTAGE SHIFT IN PRESENCE OF INTERFACE TRAPS IN JLFETS}
\begin{figure*}[t]
	\centering
	\includegraphics[width=2\columnwidth]{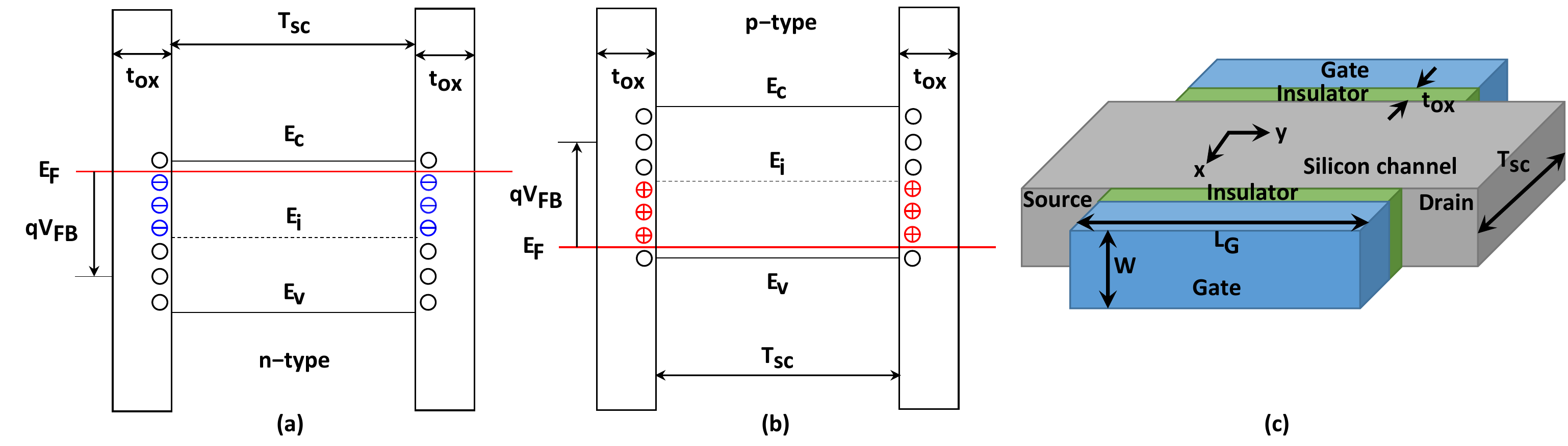}
	\caption{Energy diagram for a symmetric double-gate (a) n-type JLFET and (b) p-type JLFET at flat band voltage. (c) 3-D Schematic view of a double-gate JLFET. }
	\label{Fig1}\vspace{-0.5cm}
\end{figure*}
As mentioned, high-energy radiation and aging  are responsible for ionization damage in the form of oxide charge traps ($ Q_{ot} $) and interface charge traps ($ Q_{it} $). Traps close to the Si/oxide interface can be charged positively, negatively or neutral. These charge states depend on their energy level respect to the mid-gap energy and Fermi level. When the energy level of the trap is above the mid-gap, it behaves as an acceptor-like trap, and when below the mid-gap, it behaves like a donor-like trap \cite{jazaeri_sallese_2018} and \cite{doi:10.1002/9780470068328.ch4}. A donor-like trap with an energy level above the Fermi level will be positively charged by emitting an electron, and neutral if its energy is below the Fermi level. Conversely, an acceptor-like trap with an energy level above the Fermi level is electrically neutral and will be negatively charged by trapping an electron if its energy is below the Fermi level \cite{yesayan2016charge} and \cite{jazaeri_sallese_2018}. Hence, Interface charge traps are amphoteric and their behavior as donors or acceptors depends on their energy in the band-gap \cite{yesayan2016charge}, \cite{jazaeri2017charge}. Fig. \ref{Fig1}.a shows the energy diagram of an n-type DG JLFET at flat band voltage. As explained, the donor-like traps are completely filled but the acceptor-like traps are partially filled. Hence, we have negative charges at the Si/oxide interface. Fig. \ref{Fig1}.b demonstrates the energy diagram of a p-type DG JLFET at flat band voltage. Acceptor-like traps are completely empty and donor-like traps are partially filled, resulting in positive charges at the interface. 

The interface charge density $ Q_{it} $ is obtained by integrating the trap-state density $ D{it}(E) $ times the trap occupation probability $ f(E) $ over the band-gap energy \cite{doi:10.1002/9780470068328.ch4}, 

\begin{equation} \label {1}
	Q_{it} = -q\int\limits_{E_V}^{E_C} {D_{it}(E)}{\times}{f(E)} dE.
\end{equation}
We consider an n-type long-channel symmetric double-gate JLFET (as shown in Fig. \ref{Fig1}.c) with a doping density $  N_D $, a channel width and length $ W $ and $ L_g $  respectively, a channel thickness $ T_{sc} $ and a gate oxide thickness $ t_{ox} $.

The total charge in the semiconductor, $ Q_{sc} $, is related to the potential which drops across the gate capacitor
\begin{equation} \label {6}
Q_{sc}=-2C_{ox}\left(V_{GS}-\Delta\phi_{ms}-\psi_{s}+\frac{Q_{T}}{C_{ox}}\right),
\end{equation}
where $ \Delta\phi_{ms} $ denotes the difference between the work function of metal and the work function of an intrinsic semicondoctor, and $ Q_{T} $ represents the trap charge density includes interface charge traps and oxide charged traps i.e., $ Q_T=Q_{it}+Q_{ot} $. We rewrite (\ref{6}) as follows
\begin{equation} \label {7}
V_{GS}=\Delta\phi_{ms}+\psi_{s}-\frac{Q_{sc}+2Q_{T}}{2C_{ox}}.
\end{equation}
The surface potential only depends on the charge density in the silicon channel ($ Q_{sc}$). Hence, by imposing the same charge in the semiconductor, the gate potential in the case there is no trap i.e., $ Q_{T}=0 $ will be
\begin{equation} \label {8}
V'_{GS}=\Delta\phi_{ms}+\psi_{s}-\frac{Q_{sc}}{2C_{ox}}.
\end{equation}
Therefore, according to (\ref{7}) and (\ref{8}), traps are responsible for a gate-source voltage shift ($ \Delta V_{GS} $) equal to $ -Q_T/C_{ox} $. As expected, two contributions are evidenced
\begin{equation} \label {9}
\Delta V_{GS}=\frac{-Q_{it}}{C_{ox}}+\frac{-Q_{ot}}{C_{ox}}.
\end{equation}
The oxide charge traps are supposed to have a fixed charge density (i.e. they are not affected by the potentials) meaning that they would merely shift the electrical characteristics by a constant value $ -Q_{ot}/C_{ox}$. Therefore, without loss of generality, we propose to focus on the effect of $Q_{it}$ on the electrical characteristics.

Because the continuum nature of traps energy levels, (\ref{1}) has no analytical solution. Thus, we first investigate the single-level interface traps, then include the common exponential trap energy level distribution.
\section{Single Energy Level Traps} \label{sec:singletrap}
We assume an n-type DG JLFET and consider acceptor-like interface traps with a density ($ N_{it} $) per unit area at $ E_{it} $ which is the energy of single interface charge trap. Hence, for a single trap $ (E_{it}) $, $ D_{it}(E_{it}) $ equals to $ -q \times N_{it} $ and $ Q_{it} $ is given by $ Q_{it}=-q\times N_{it} \times f(E_{it}) $. The occupation probability of an acceptor-like trap with an energy level $ E_{it} $ is determined by \cite{jazaeri2017charge}
\begin{equation} \label {2}
f(E_{it})^{-1} = 1+\exp\left(\frac{E_{it}-E_F}{k_{B}T}\right).
\end{equation}
Here we assume the degeneracy factor equal to 1. By defining $ E_{t-i}=E_{it}-E_{i} $, where $ E_{i} $ is the intrinsic fermi level, $ Q_{it} $ is obtained by
\begin{equation} \label {21}
\frac{Q_{it}}{qN_{it}} =-\left[1+\exp\left(\frac{E_{t-i}}{k_{B}T}\right)\exp\left({-\frac{\psi_{s}-V_{ch}}{U_{T}}}\right)\right]^{-1},
\end{equation} 
where $ \psi_{s}=-(E_{i}-E_{f})/q $ is the surface potential, $ V_{ch} $ is the quasi Fermi potential and $ U_{T}=k_BT/q $ is the thermal voltage.

Following the derivation of the charge-based model for double-gate JLFETs developed in \cite{sallese2011charge} and \cite{jazaeri_sallese_2018}, we have the two following relationships which link the surface potential ($ \psi_{s} $) and the center potential ($ \psi_{0} $)
\begin{equation} \label {3}
\begin{split}
E_{s}^2&=\frac{2qn_{i}U_{T}}{\epsilon_{si}}\Biggl[\exp\left({\frac{\psi_{s}-V_{ch}}{U_{T}}}\right)-\exp\left({\frac{\psi_{0}-V_{ch}}{U_{T}}}\right)\\
&-\frac{N_D}{n_i}\left(\frac{\psi_{s}-\psi_{0}}{U_{T}}\right)\Biggr],
\end{split}
\end{equation}
\begin{equation} \label {31}
\psi_{s}-\psi_{0}=\frac{qn_{i}T_{sc}^2}{8\epsilon_{si}}\Biggl[\exp\left({\frac{\psi_{0}-V_{ch}}{U_{T}}}\right)-\frac{N_D}{n_i}\Biggr],
\end{equation}
where $ n_i $ is the intrinsic carrier concentration and $ \epsilon_{si} $ is the permittivity of silicon. $ E_{s} $ is the surface electric field which equals to $ {Q_{sc}}/{2\epsilon_{si}} $. Hence, $ Q_{sc} $, the semiconductor total charge density can be linked to the surface and center potential, and we can write
\begin{equation} \label {4}
\begin{split}
\left(\frac{Q_{sc}}{2\epsilon_{si}}\right)^2&\!\!\!=\frac{2qn_{i}U_{T}}{\epsilon_{si}}\Biggl\{\!\exp\left({\frac{\psi_{0}-V_{ch}}{U_{T}}}\right)\!\!\left[\exp\left({\frac{\psi_{s}-\psi_{0}}{U_{T}}}\right)\!-1\right]\\
&-\frac{N_D}{n_i}\left(\frac{\psi_{s}-\psi_{0}}{U_{T}}\right)\Biggr\}.
\end{split}
\end{equation}
\vspace{-1cm}
\subsection{Depletion Mode}
Under depletion mode, the potential at the center is larger than the potential at the surface which leads to a positive charge in the semiconductor ($ Q_{sc}\geq0 $). Hence, the exponential term in (\ref{4}) is much smaller than the second term and therefore relation (\ref{4}) can be approximated with
\begin{equation} \label {41}
\left(\frac{Q_{sc}}{2\epsilon_{si}}\right)^2=\frac{2qN_D}{\epsilon_{si}}\left(\psi_{0}-\psi_{s}\right).
\end{equation}
By substituting (\ref{31}) in (\ref{41}), the surface potential in the depletion mode is obtained with respect to the total charge density
\begin{equation} \label {5}
\psi_{s}-V_{ch}=U_{T}\ln\left(\alpha\theta\right),
\end{equation}
where $ \alpha$ and $ \theta$ are
\begin{equation} \label {51}
\alpha=\frac{N_D}{n_i}\exp\left(-\frac{Q_{sc}^2}{8U_TC_{si}Q_{f}}\right),
\end{equation}
\begin{equation} \label {52}
\theta=1-\left(\frac{Q_{sc}}{Q_{f}}\right)^2,
\end{equation}
where $ Q_f=qN_DT_{si} $ and $ C_{si}=\epsilon_{si}/T_{sc} $.

Unlike oxide charge traps, interface charge traps can cause a gate-source voltage shift that depends on the total charge density. By introducing (\ref{5}) in (\ref{21}) and then in (\ref{9}) (with $ Q_ {ot}=0 $), the gate-source voltage shift is
\begin{equation} \label {10}
\Delta V_{GS} = \frac{qN_{it}}{C_{ox}\left(1+\eta \: \alpha^{-1}\theta^{-1}\right)},
\end{equation}
where $ \eta=\exp(E_{t-i}/k_BT) $. Hence, the total charge density in the semiconductor becomes
\begin{equation} \label {11}
Q_{sc}=-2C_{ox}\left(V_{GS}^*-\Delta V_{GS}-V_{ch}-U_{T}\ln\left(\alpha\theta\right)\right),
\end{equation}
where we define the effective gate voltage as $ V_{GS}^*=V_{GS}-\Delta\phi_{ms} $. By solving relationships (\ref{10}) and (\ref{11}), total charge density in the semiconductor and the mobile charge density $ Q_m=Q_{sc}-Q_{f} $, are obtained. It is worth mentioning that, when there are not traps i.e., $ Q_ {T}=0 $, $ \Delta V_{GS} $ is zero and relationship (\ref{11}) gives back the general relationship derived in \cite{sallese2011charge} and \cite{jazaeri_sallese_2018}.
\subsection{Threshold Voltage Shift} 
Unlike inversion mode MOSFETs, the definition of threshold voltage ($ V_{th} $) in junctionless transistors is not obvious. However,  there are some definitions of threshold voltage for junctionless transistors. One of these defines the threshold voltage as the gate potential which cancels the mobile charge density ($Q_{sc}=Q_f$) when the logarithmic term in (\ref{11}), i.e. $ \theta $ is neglected \cite{jazaeri_sallese_2018}. Hence, $ \alpha $ at threshold takes a simple form
\begin{equation} \label {53}
\alpha=\frac{N_D}{n_i}\exp\left(-\frac{Q_{f}}{8U_TC_{si}}\right).
\end{equation}
Under these assumptions, the threshold voltage shift is
\begin{equation} \label {101}
\Delta V_{th} = \frac{qN_{it}}{C_{ox}(1+\eta \: \alpha^{-1})}.
\end{equation}
\vspace{-0.8cm}
\subsection{Accumulation Mode}
In accumulation mode, the second term in (5) is always smaller than the exponential term. Therefore, the second term in accumulation can be omitted. In addition, in accumulation the center potential remains close to the value it takes at the flat-band condition \cite{jazaeri_sallese_2018}
\begin{equation} \label {32}
\psi_{0}\approx\psi_{0_{FB}}=V_{ch}+U_T\ln\left(\frac{N_D}{n_i}\right).
\end{equation}
Thus, relation (\ref{4}) can be simplified to
\begin{equation} \label {42}
\left(\frac{Q_{sc}}{2\epsilon_{si}}\right)^2=\frac{2qn_iU_T}{\epsilon_{si}}\left[\exp\left({\frac{\psi_{s}-V_{ch}}{U_{T}}}\right)-\frac{N_D}{n_i}\right].
\end{equation}
Hence, the surface potential in the accumulation mode is derived and is a function of the total charge density 
\begin{equation} \label {12}
\psi_{s}-V_{ch}=U_{T}\ln\left(\beta\right),
\end{equation}
where $ \beta $ is defined as
\begin{equation} \label {121}
\beta=\frac{Q_{sc}^2}{8\epsilon_{si}qn_iU_T}+\frac{N_D}{n_i}.
\end{equation}
Introducing (\ref{12}) in (\ref{21}) and then in (\ref{9}) (with $ Q_ {ot}=0 $) results in the gate-source voltage shift in accumulation mode
\begin{equation} \label {72}
\Delta V_{GS} = \frac{qN_{it}}{C_{ox}\left(1+\eta\beta^{-1}\right)},
\end{equation}
and the total charge density in the semiconductor in accumulation mode becomes
\begin{equation} \label {71}
Q_{sc}=-2C_{ox}\left(V_{GS}^*-\Delta V_{GS}-V_{ch}-U_{T}\ln\left(\beta\right)\right).
\end{equation}
Comparing (\ref{10}) with (\ref{72}), we notice that $ \Delta V_{GS} $ is given by the same relationship for both regions of operation, an unexpected but very interesting result which can be summarized as follows
\begin{equation} \label {13}
\Delta V_{GS} = \frac{qN_{it}}{C_{ox}\left(1+\eta \: \gamma^{-1}\right)}, \:\:\gamma=\begin{cases} \alpha\theta, & \mbox{Depletion} \\ \beta, & \mbox{Accumulation} \end{cases}.
\end{equation}
This implies that the mobile charge density satisfies a generic relationship,
\begin{figure*}[t]
	\centering
	\includegraphics[width=2\columnwidth]{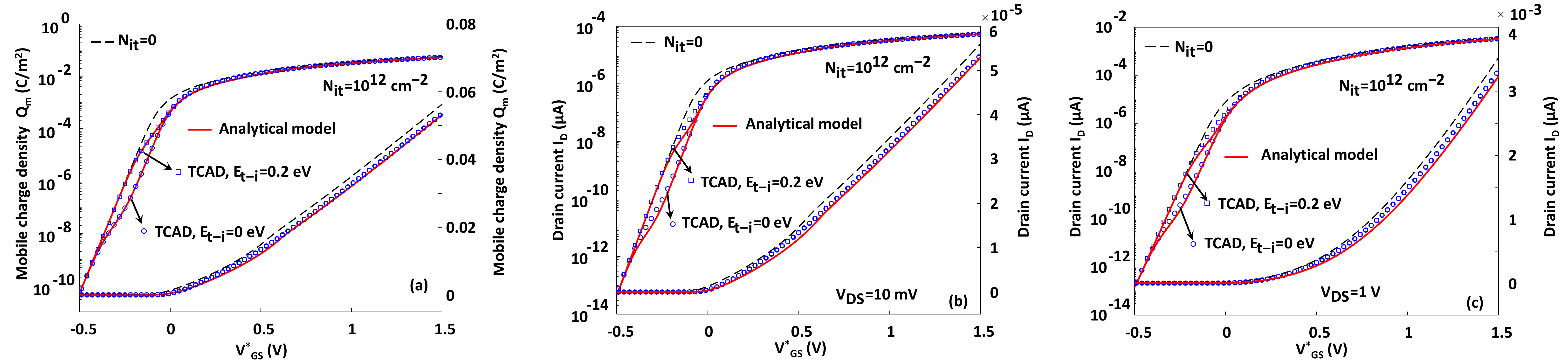}
	\caption{(a) Mobile chareg density versus effective gate voltage calculated from analytical model and TCAD simulations in logaritmic (left axis) and linear (right axis) scale for traps with energy levels $ E_{t-i}$ = \SI{0}{eV} and $ E_{t-i}$ = \SI{0.2}{eV}. Drain current versus effective gate voltage calculated from analytical model and TCAD simulations in logaritmic (left axis) and linear (right axis) scale for traps with energy levels $ E_{t-i}$ = \SI{0}{eV} and $ E_{t-i}$ = \SI{0.2}{eV} (b) at $ V_{DS}$ = \SI{10}{mV} and (c) $ V_{DS}$ = \SI{1}{V}.}
	\label{Fig2}
\end{figure*}
\begin{figure*}[t]
	\centering
	\includegraphics[width=2\columnwidth]{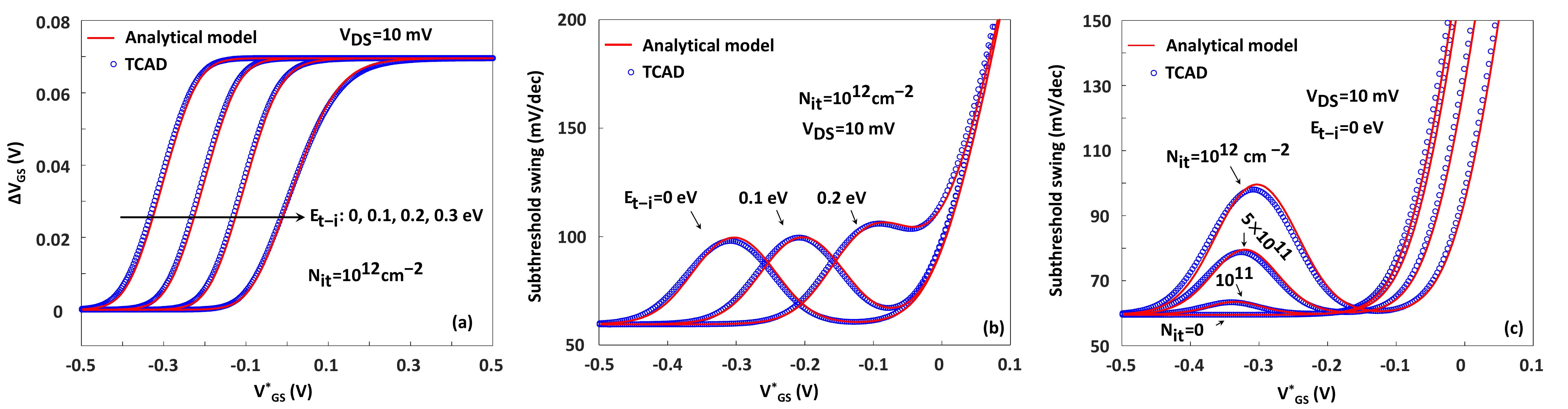}
	\caption{ (a) Gate voltage shift versus effective gate voltage calculated from analytical model and TCAD simulations for traps with energy levels $ E_{t-i}$ = \SI{0}{eV}, $ E_{t-i}$ = \SI{0.1}{eV}, $ E_{t-i}$ = \SI{0.2}{eV} and $ E_{t-i}$ = \SI{0.3}{eV}. (b) Subthreshold swing versus effective gate voltage calculated from analytical model and TCAD simulations for traps with energy levels $ E_{t-i}$ = \SI{0}{eV}, $ E_{t-i}$ = \SI{0.1}{eV}, and $ E_{t-i}$ = \SI{0.2}{eV}. (c) Subthreshold swing versus effective gate voltage calculated from analytical model and TCAD simulations for $ N_{it}$ = \SI{0}{cm^{-2}}, $ N_{it}$ = \SI{1e11}{cm^{-2}}, $ N_{it}$ = \SI{5e11}{cm^{-2}}, and $ N_{it}$ = \SI{1e12}{cm^{-2}}.}
	\label{Fig3}\vspace{-0.5cm}
\end{figure*}
\begin{equation} \label {131}
Q_{m}=-Q_f-2C_{ox}\left(V_{GS}^*-\Delta V_{GS}-V_{ch}-U_{T}\ln(\gamma)\right).
\end{equation}
To confirm the validity of this model, we performed simulations with SILVACO TCAD software assuming a double gate JLFET with \SI{1}{\micro m} channel length and \SI{10}{nm} silicon thickness. The doping density and oxide thickness were set respectively to $ N_D$ = \SI{5E16}{cm^{-3}} and \SI{1.5}{nm}.

Here the single energy level has been activated for the traps. The interface trap density is set to $ N_{it}$ = \SI{1E12}{cm^{-2}}. The mobile charge density versus effective gate voltage which is obtained from TCAD simulations and the model is depicted in Fig. \ref{Fig2}.a. Lines and symbols have been used for the analytical model and TCAD simulations, respectively. Both linear and logarithmic representations of the analytical model demonstrate a full agreement with TCAD simulations.
\subsection{Drain Current}
Since we have analytical relationships for mobile charge density in depletion and accumulation modes, we can use both analytical charge-based model described in \cite{sallese2011charge} and explicit relationships which are derived in \cite{yesayan2013explicit} for obtaining drain current. To assess this model with TCAD simulations, we have used explicit relationships and we recall them here. Thus for the drain current in the depletion mode we have the following equation from \cite{yesayan2013explicit}
\begin{equation} \label {22}
\begin{split}
&I_{Dep}\left(V^*_G, V_D, V_S\right)\!=\!\mu\frac{W}{L_G}\Biggl[\!\left(\frac{1}{8C_{si}}-\frac{1}{4C_{ox}}\right)\!Q_{sc}^2\!-\!\frac{Q_{sc}^3}{12Q_fC_{si}}\\
&+\left(\frac{Q_f}{2C_{ox}}+2U_T\right)Q_{sc}-U_TQ_f\ln\left(1+\frac{Q_{sc}}{Q_f}\right)^2\Biggr]^D_S,
\end{split}
\end{equation}
and for the accumulation mode \cite{yesayan2013explicit}
\begin{equation} \label {23}
\begin{split}
&I_{Acc}\left(V^*_G, V_D, V_S\right)\!=\!\mu\frac{W}{L_G}\Biggl[\left(\frac{Q_f}{2C_{ox}}+2U_T\right)Q_{sc}-\frac{1}{4C_{ox}}Q_{sc}^2\\
&-U_TQ_f\ln\left(1+\frac{Q_{sc}^2}{8Q_fC_{si}U_T}\right)\\
&-2U_T\sqrt{8Q_fC_{si}U_T}\arctan\left(\frac{Q_{sc}}{\sqrt{8Q_fC_{si}U_T}}\right)\Biggr]^D_S,
\end{split}
\end{equation}
where $ \mu$ is the free carrier mobility. If we define the hybrid channel when some part of the channel, near the source contact, is in accumulation and the other part, near the drain, is in depletion, we can write the drain current as follows \cite{yesayan2013explicit}

\begin{equation} \label {24}
\begin{split}
&I_{hyb}\left(V^*_G, V_D, V_S\right)=I_{Acc}\left(V^*_G, V^*_G-V_{FB}, V_S\right)\\
&+I_{Dep}\left(V^*_G, V_D, V^*_G-V_{FB}\right).
\end{split}
\end{equation}
In this work $ \mu $, is assumed constant along the channel and the width of the device is set to \SI{1}{\mu m}.

By solving (\ref{131}) and intruducing it in (\ref{22}) to (\ref{24}) the drain current is obtained. The drain current versus the effective gate voltage at $ V_{DS}$ = \SI{10}{mV} and $ V_{DS}$ = \SI{1}{V} for the trap energy levels $ E_{t-i}$ = \SI{0}{eV} and $ E_{t-i}$ = \SI{0.2}{eV} has been plotted in Fig. \ref{Fig2}.b and \ref{Fig2}.c. The results from the model are compared with TCAD simulations using the same parameters and indicate a good agreement in both linear and exponential representations.

The black dashed line in all the figures depicts the case which the density of traps is zero i.e. without traps. In order to make the effect of interface traps more clear, gate voltage shift due to the interface traps as a function of effective gate voltage for different values of energy levels has been shown in Fig. \ref{Fig3}.a. The agreement between the analytical model and TCAD simulations is excellent in both depletion and accumulation modes. It is observed from the plot that $ \Delta V_{GS}$ for above the flatband voltage doesn't depend on the $ E_{t-i}$ and it can be simplified to $ \Delta V_{GS}=qN_{it}/C_{ox}$.

\subsection{Subthreshold Swing}
\begin{figure}[t]
	\centering
	\includegraphics[width=1\columnwidth]{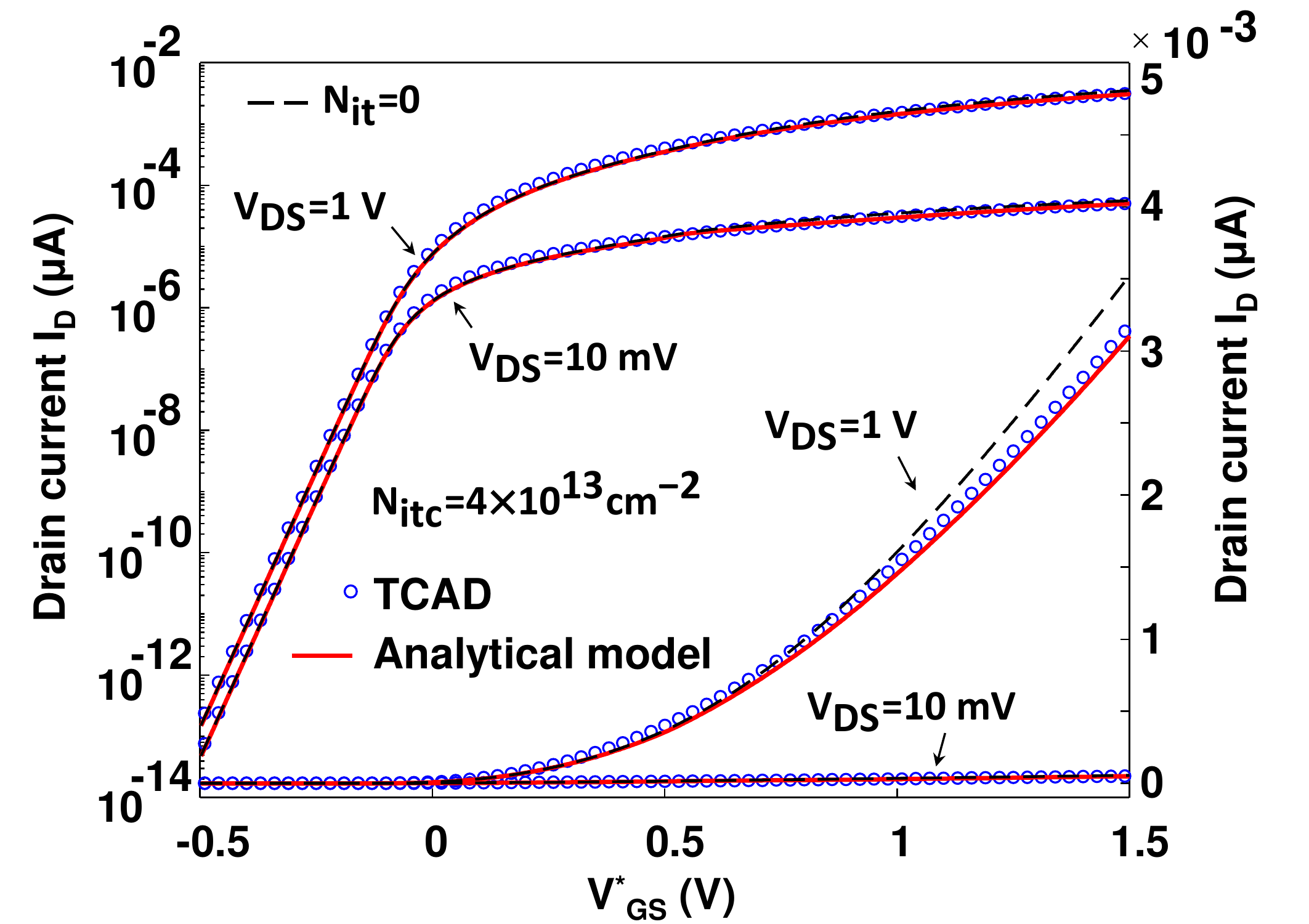}
	\caption{Drain current versus effective gate voltage calculated from analytical model and TCAD simulations in logaritmic (left axis) and linear (right axis) scale for traps with exponential energy level distribution at $ V_{DS}$ = \SI{10}{mV} and $ V_{DS}$ = \SI{1}{V}.}
	\label{Fig4}\vspace{-0.5cm}
\end{figure}
Subthreshold swing degradation is another important effect of interface charge traps since they implicitly change the charge density with voltages. In subthreshold regime, saturation is reached at low drain-to-source voltages (about $ 4U_T$). By assuming that the mobile charge density at low $ V_{DS} $ is almost constant along the channel from source to drain, the current can be approximated by
\begin{equation} \label {151}
I_D=\frac{W}{L_g} \mu Q_m V_{DS},
\end{equation}
and the subthreshold swing ($ SS$) becomes
\begin{equation} \label {15}
SS=\left[\frac{\partial }{\partial V_{GS}} \log\left(I_D\right)\right]^{-1}=Q_m \ln\left(10\right) \frac{\partial V_{GS}}{\partial Q_m},
\end{equation}
where $ {\partial V_{GS}}/{\partial Q_m} $ derived from (\ref{131}):
\begin{equation} \label {16}
\frac{\partial V_{GS}}{\partial Q_m}=-\frac{1}{2C_{ox}}+\frac{\gamma\sp{\prime}}{\gamma}\left(\frac{qN_{it}\:\eta\:\gamma}{C_{ox}\left(\eta+\gamma\right)^2}+U_T\right),
\end{equation}
where $ \gamma\sp{\prime}=\partial \gamma/\partial Q_m$. Since the device operates in subthreshold, according to relation (\ref{13})  $ \gamma $ is equal to $ \alpha\theta$ and $ \gamma\sp{\prime}/\gamma$ becomes
\begin{equation} \label {17}
\frac{\gamma\sp{\prime}}{\gamma}=\frac{2Q_{sc}}{Q_{m}\left(Q_{m}+2Q_{f}\right)}-\frac{Q_{sc}}{4U_TC_{si}Q_{f}}.
\end{equation}
Next, introducing (\ref{17}) in (\ref{16}) then substituting in (\ref{15}), the subthreshold swing becomes an explicit function of the mobile charge density:
\begin{equation} \label {18}
SS\!=\!\ln\left(10\right)\!\!\left[-\frac{Q_m}{2C_{ox}}\!+\!\frac{Q_m \gamma\sp{\prime}}{\gamma}\!\!\left(\frac{qN_{it}\:\eta\:\gamma}{C_{ox}\left(\eta+\gamma\right)^2}+U_T\right)\!\right]\!\!.
\end{equation}
To extract the maximum subthreshold swing ($ SS_{max} $), we propose to introduce some approximation in (\ref{18}). Since $ Q_m $ is negligible in the subthreshold, we can ignore $ Q_m/(2Cox)$ and assume that $ Q_m\gamma\sp{\prime}/\gamma$ is close to unity:
\begin{equation} \label {19}
SS\approx\ln\left(10\right)\left[\frac{qN_{it}\:\eta\:\gamma}{C_{ox}\left(\eta+\gamma\right)^2}+U_T\right].
\end{equation}
Thus, $ SS $ peaks when $ \gamma=\eta $
\begin{equation} \label {40}
SS_{max}=\ln\left(10\right)\left(\frac{qN_{it}}{4C_{ox}}+U_T\right).
\end{equation}
Interestingly, the largest value of the subthreshold slope is closely related to the density of interface traps.

Subthreshold swing degradation induced by the presence of interface traps can be observed in Fig. \ref{Fig2}. To make more clear this effect, subthreshold swing as a function of the effective gate potential for different values of  $ E_{t-i}$ and  $ N_{it}$ for both the analytical model and TCAD simulations, have been plotted in Fig. \ref{Fig3}.a and \ref{Fig3}.b. The agreement between the analytical solution and the TCAD simulations is evidenced.

\section{Exponential Trap Energy Level Distribution}
So far only considered discrete energy trap levels have been discussed. Although these are instructive, they are not representative of real devices. Admittedly, real devices exhibit an exponential distribution of traps energy in the bandgap, commonly designed as U-shaped distribution \cite{kim2003distribution}.

Since we assume an n-type device, only interface traps which energies above the midgap are considered. In addition, according to the TCAD models, the maximum in the acceptor-like density lays at the conduction band edge. Hence, the charge density of traps can be written
\begin{equation} \label {100}
Q_{it} = -q\int\limits_{E_i}^{E_C} {N_{itc}\exp\left(\frac{E-E_C}{E_d}\right)}{\times}{f(E)} dE,
\end{equation}
where $ N_{itc}$ is the density of acceptor-like states in the exponential distribution at the conduction band edge and $ E_d$ specifies the characteristic decay energy.

To calculate the trapped charge density, we rely on the same model discussed in section \ref{sec:singletrap} valid for for single trap energy levels where we simply propose to replace $ N_{it}$ and $ E_{t-i}$ with averaged parameters $ N_{it}^*$ and $ E_{t-i}^*$:
\begin{equation} \label {212}
\frac{Q_{it}}{qN_{it}^*} = -\left[1+\exp\left(\frac{E_{t-i}^*}{KT}\right)\gamma^{-1}\right]^{-1}.
\end{equation} 
The analytical approach is compared with TCAD simulations as shown in Fig. \ref{Fig4}. Using $ N_{itc}$ = \SI{4e13}{cm^{-2}} and $ E_{d}$ = \SI{0.035}{eV} as reported in \cite{yesayan2016charge}, Then by adjusting the averaged parameters, $ N_{it}^*$ = \SI{1.5e12}{cm^{-2}} and $ E_{t-i}^*$ = \SI{0.592}{eV}. These values fit quite well with TCAD results. Interestingly, these parameters which were 'extracted' at room temperature still give accurate results when changing the temperature. Fig. \ref{Fig4} illustrates the drain current versus the effective gate potential at $ V_{DS}$ = \SI{10}{mV} and $ V_{DS}$ = \SI{1}{V} for an exponential distribution of interface traps energy. Since the maximum density of interface traps happens close to the conduction band edge, their impact on the electrical characteristic is evidenced at relatively high gate potentials, otherwise traps remain unoccupied.

\section{Assessment of the Model from 77K to 400K}
\begin{figure*}[t]
	\centering
	\includegraphics[width=2\columnwidth]{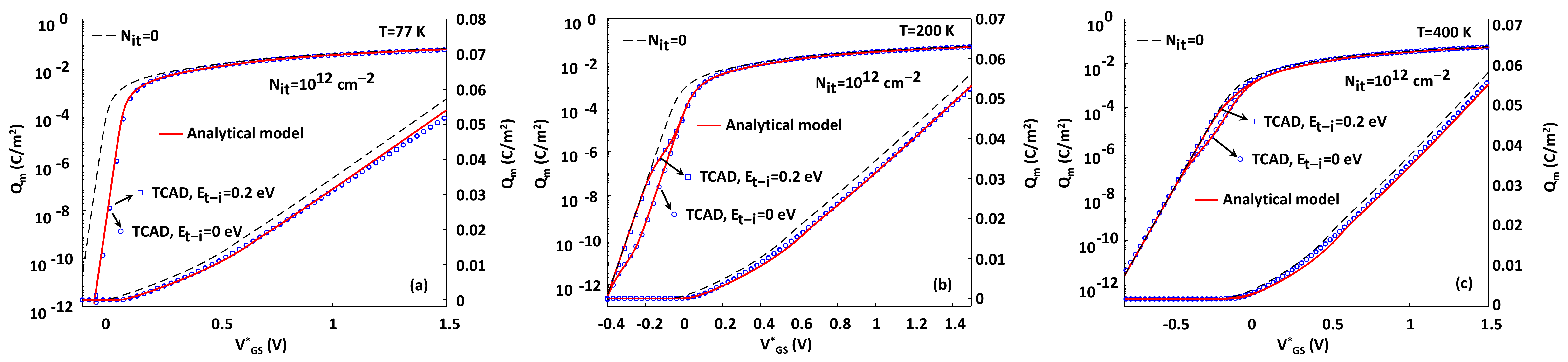}
	\caption{Mobile charge density versus effective gate voltage calculated from analytical model and TCAD simulations in logaritmic (left axis) and linear (right axis) scale for the single trap energy levels $ E_{t-i}$ = \SI{0}{eV} and $ E_{t-i}$ = \SI{0.2}{eV} at (a) $ T$ = \SI{77}{K}, (b) \SI{200}{K}, and (c) \SI{400}{K}.}
	\label{Fig5}
\end{figure*}
\begin{figure*}[h]
	\centering
	\includegraphics[width=2\columnwidth]{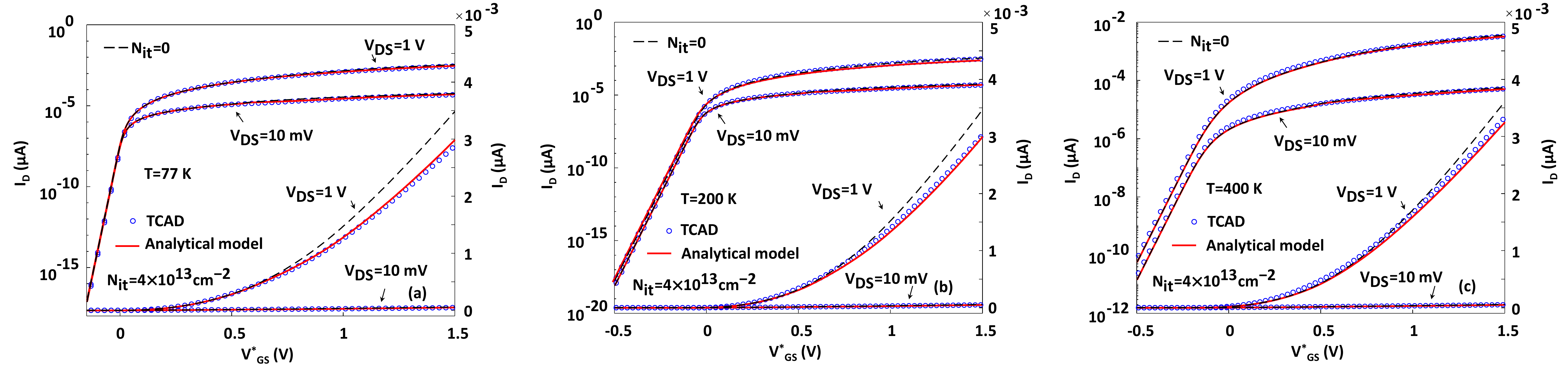}
	\caption{Drain current versus effective gate voltage calculated from analytical model and TCAD simulations in logaritmic (left axis) and linear (right axis) scale for traps with exponential energy level distribution at $ V_{DS}$ = \SI{10}{mV} and $ V_{DS}$ = \SI{1}{V} at (a) $ T$ = \SI{77}{K}, (b) \SI{200}{K}, and (c) \SI{400}{K}.}
	\label{Fig6}\vspace{-0.5cm}
\end{figure*}
In some applications, low temperature operation is necessary and present many advantages such as a steeper subthreshold slope \cite{beckers2018cryogenic, rogers1968most, beckers2018characterization}. Having an analytical model covering a wide range of temperature operation is therfore a big advantage.

In this section, we assess the model for various temperatures ranging from $ \SI{77}{K}$ to $ \SI{400}{K}$ (note that we have used Boltzmann statistics only). For simplicity, we also assume a constant mobility, given that this will mainly act as a scaling factor for the current, but will have almost no impact on the electrostatics, i.e. on the conclusions of this section \cite{jazaeri2015carrier}. According to \cite{balestra2017physics} the mobility at low temperatures changes with respect to the mobile charge density. This effect could be introduced in the proposed model at the correction to the current, however, the conclusions on the trap charge distribution will remain the same. In addition, this would require introducing fitting parameters, which would weaken our physics-based analysis. The intrinsic carrier concentration for the temperatures used in this section are $ n_{i}$ = \SI{4.39e-13}{m^{-3}}, \SI{1.18e11}{m^{-3}}, and \SI{6.16e18}{m^{-3}} at $ T$ = \SI{77}{K}, \SI{200}{K}, and \SI{400}{K} respectively. 

Fig. \ref{Fig5} depicts the mobile charge density versus the effective gate voltage at $ T$ = \SI{77}{K}, \SI{200}{K}, and \SI{400}{K} for single trap energy levels $ E_{t-i}$ = \SI{0}{eV} and $ E_{t-i}$ = \SI{0.2}{eV}. For the range of mobile charge density considered, which is depicted in Fig. \ref{Fig5}, different trap energy levels behave in the same way at \SI{77}{K}. Indeed for those energy levels the traps do not change their charge states. The results confirm an excellent agreement between the analytical model and TCAD simulations.

Finally, we also assessed the validity of the model for the more realistic case of an exponential energy trap distribution at $ T$ = \SI{77}{K}, \SI{200}{K}, and \SI{400}{K}. These are shown in Fig. \ref{Fig6} where the drain current versus the effective gate voltage is plotted at $ V_{DS}$ = \SI{10}{mV} and $ V_{DS}$ = \SI{1}{V}. Interestingly, the averaged parameters, i.e., $ N_{it}^*$ and $ E_{t-i}^*$ do not need to be modified, meaning that the model that we have presented is quite predictive (note that we anticipate that for lower temperatures i.e. \SI{4.2}{K} the roots of the model would have to be revised introducing Fermi-Dirac Statistics, and possibly 2D density of states).

\section{Conclusion}
An analytical charge-based model for symmetric double-gate junctionless FETs with interface charge traps was developed. The model incorporates the impact of radiation and aging degradation on DC electrical characteristics of double-gate JLFETs by proposing an equivalent gate-source voltage. A detailed study of the interface charge traps and their influence on the device performance is carried out. Both single energy level and exponential distribution energy levels for interface traps have been investigated. In particular, the subthreshold swing degradation in the presence of single level interface charge traps has been modelled accurately. We also included the impact of the temperature from \SI{77}{K} to \SI{400}{K}, a very important aspect for cryogenic applications. The model has been compared to TCAD simulations with an excellent agreement in all regions of operation from deep depletion to accumulation and linear to saturation.

\bibliographystyle{IEEEtran}

\end{document}